\documentclass[a4paper,floatfix,accepted=2024-06-21]{quantumarticle}
\pdfoutput=1
\usepackage{graphicx} 
\usepackage{amsmath}
\usepackage{amssymb}
\usepackage{amsthm}
\usepackage{comment}
\usepackage{physics}
\usepackage{xcolor}
\usepackage{float}
\usepackage{flushend}
\usepackage[normalem]{ulem}
\usepackage{hyperref}
\usepackage[numbers]{natbib}

\newcommand{\be}{\begin{equation}}
\newcommand{\ee}{\end{equation}}

\newcommand{\calN}{\mathcal{N}}
\newcommand{\calO}{\mathcal{O}}
\newcommand{\calH}{\mathcal{H}}
\newcommand{\calF}{\mathcal{F}}
\newcommand{\calL}{\mathcal{L}}
\newcommand{\calG}{\mathcal{G}}
\newcommand{\calI}{\mathcal{I}}

\newcommand{\dinv}{\delta^{-1}}

\newcommand{\sumP}{\left ( \sum_{i=1}^N P_i \right )}
\newcommand{\sumPsq}{\left ( \sum_{i=1}^N P^2_i \right )}
\newcommand{\poly}{\text{poly}}
\newcommand{\varL}{\sigma^2_{\calL}}

\newcommand{\Id}{\mathbb{I}}
\newcommand{\supp}{\text{sup }}

\begin{document}

\title{Efficient Quantum Algorithm for Filtering Product States}


\author{Reinis Irmejs}
\email{reinis.irmejs@mpq.mpg.de}
\affiliation{Max-Planck-Institut für Quantenoptik, Hans-Kopfermann-Straße 1, D-85748 Garching, Germany}
\affiliation{Munich Center for Quantum Science and Technology (MCQST), Schellingstraße 4, D-80799 Munich, Germany}

\author{Mari Carmen Bañuls}
\affiliation{Max-Planck-Institut für Quantenoptik, Hans-Kopfermann-Straße 1, D-85748 Garching, Germany}
\affiliation{Munich Center for Quantum Science and Technology (MCQST), Schellingstraße 4, D-80799 Munich, Germany}

\author{J. Ignacio Cirac}
\affiliation{Max-Planck-Institut für Quantenoptik, Hans-Kopfermann-Straße 1, D-85748 Garching, Germany}
\affiliation{Munich Center for Quantum Science and Technology (MCQST), Schellingstraße 4, D-80799 Munich, Germany}

\begin{abstract}
    We introduce a quantum algorithm to efficiently prepare states with a small energy variance at the target energy. We achieve it by filtering a product state at the given energy with a Lorentzian filter of width $\delta$. Given a local Hamiltonian on $N$ qubits, we construct a parent Hamiltonian whose ground state corresponds to the filtered product state with variable energy variance proportional to $\delta\sqrt{N}$. We prove that the parent Hamiltonian is gapped and its ground state can be efficiently implemented in $\poly(N,1/\delta)$ time via adiabatic evolution. We numerically benchmark the algorithm for a particular non-integrable model and find that the adiabatic evolution time to prepare the filtered state with a width $\delta$ is independent of the system size $N$. Furthermore, the adiabatic evolution can be implemented with circuit depth $\mathcal{O}(N^2\delta^{-4})$. Our algorithm provides a way to study the finite energy regime of many body systems in quantum simulators by directly preparing a finite energy state, providing access to an approximation of the microcanonical properties at an arbitrary energy.
\end{abstract}

\maketitle
\section{Introduction}

Quantum computing introduces a novel approach to computational tasks, with one of the main advantages being the efficient simulation of physical quantum systems. The advantage lies in the potential for exponential enhancement in spatial resources by encoding the system's degrees of freedom into qubits, enabling direct operations and calculations on them \cite{deutsch1985quantum, lloyd1996universal, cirac2004new, QSimReview}.

A particularly compelling application pertains to determining the ground state properties of systems – a realm of zero-temperature behavior. While, in general, identifying the ground state of a local Hamiltonian stands as a QMA-hard problem \cite{Kempe2006,cubitt2016complexity}, for typical physically relevant systems solutions are attainable. Quantum adiabatic evolution offers a method for preparing the ground state of a specified Hamiltonian. This process commences from a readily preparable ground state of the Hamiltonian $H_0$ and involves time-evolving the system to a target Hamiltonian $H_1$. This strategy's success relies on an energy gap that scales favorably with system size \cite{Farhi2000}.

Nonetheless, studying excited states is equally important, particularly to understand systems in thermal equilibrium. In practice, addressing a particular eigenstate of the system is extremely difficult. However, analyzing states of a small energy variance around the target energy is equally interesting. Having access to finite energy states of a small energy variance could be useful for studying physics in many fields.
In the field of many-body physics, one can calculate the expectation values for observables, perform time dynamics simulations, and calculate entanglement properties of the state, i.e., Renyi entropies \cite{johri2017entanglement, wang2023quantum}. Furthermore, given access to two states at different energies, it becomes possible to verify the predictions of the eigenstate thermalization hypothesis \cite{Deutsch_2018}. In chemistry, the excited states can be used to simulate molecular dynamics \cite{Sokolov2021}. 
The preparation of such states has been well studied using tensor networks \cite{Yu2017, Banuls2020, Yang2022}. In \cite{Banuls2020}, a relationship between the energy variance and the associated entanglement entropy of the state was established, showing the need for a large bond dimension to approximate the states with small energy variance.

The difficulty of analyzing the excited states classically has sparked an interest in developing quantum algorithms for this purpose. The first quantum algorithm to prepare excited states with reduced variance was quantum phase estimation (QPE) \cite{kitaev1997quantum, Abrams1999}; however, this approach is very costly. An alternative is to use time series \cite{o2019quantum, Cruz_2020, Jensen_2020, Seki_2022, nisqQPE}, which does not require preparation of the state to extract its properties; nevertheless, it uses costly controlled evolution operations. 

An alternative to the direct preparation of the state was proposed in \cite{Lu2021}, where observables at a given energy were calculated by means of a cosine filter on energy, deconstructed as a sum of Loschmidt echoes and combined with subsequent classical post-processing.

In this work, we propose a very different way of directly preparing a state of small energy variance. For a given Hamiltonian $H$, we find a product state $\ket{\Psi}$ at the target energy and filter it to a small energy variance. This is achieved by defining a parent Hamiltonian $\calH$ that depends on $H$, $\ket{\Psi}$, and on a filter width parameter $\delta$ that is proportional to the desired energy variance. By construction, the unique ground state $\ket{\Phi}$ of the parent Hamiltonian $\calH$ corresponds to the filtered product state. Furthermore, we define a gapped adiabatic path that connects the initial product state $\ket{\Psi}$ with the filtered state $\ket{\Phi}$. Thus, performing adiabatic evolution along this path allows one to efficiently prepare a finite variance approximation of the excited eigenstate at a given energy. 
This parent Hamiltonian construction is inspired by the one introduced in \cite{Ge2016} to prepare a Gibbs state of a Hamiltonian with commuting terms.

The rest of the paper is organized as follows. In Section \ref{sec:Method}, we formally introduce the algorithm, in Section \ref{sec:Results}, we perform a numerical investigation to establish tighter bounds to the algorithmic runtime, followed by a summary of our results in Section \ref{sec:Discussion} and our conclusions and outlook in Section \ref{sec:Conclusion}.

\section{Method}\label{sec:Method}

\subsection{Outline}

\begin{figure}
    \centering
    \includegraphics[width = \linewidth]{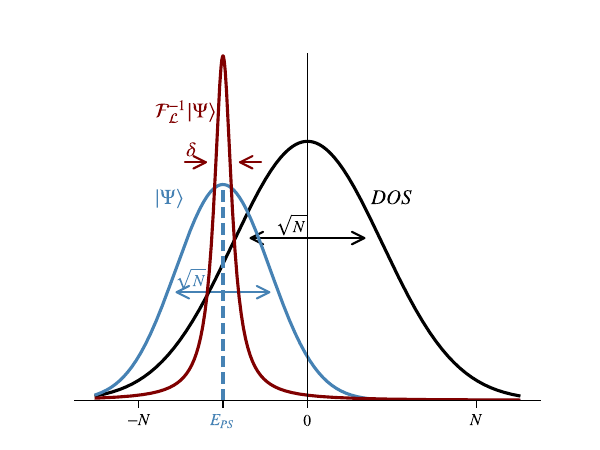}
    \caption{Illustration of the filtering. The figure shows the density of states (DOS) for the Hamiltonian $H$ and the product state $\ket{\Psi}$. The red line shows the process of applying the Lorentzian filter $\calF^{-1}_{\calL}$ onto the product state $\ket{\Psi}$, which leads to a state of reduced energy variance.}
    \label{FilteringFig}
\end{figure}

In this work, we set out to perform the following task. We start with a local Hamiltonian $H$ of interest and a product state $\ket{\Psi}$ that has energy $E_{0}$ and variance $\sigma^2_{0}$ with respect to $H$. We want to create a new state $\ket{\Phi}$ that has the same (or close) energy $E_{0}$ but a substantially reduced variance $\varL$. This is similar to preparing a microcanonical superposition at energy $E_{0}$, just with a finite width. We develop an algorithm that allows us to arbitrarily reduce the variance $\varL$ in $\poly(N, \dinv)$ time. We achieve this by applying a Lorentzian filter of width $\delta$ on top of the product state $\ket{\Psi}$ (Fig.~ \ref{FilteringFig}) and suppressing the components of eigenstates with energy far away from $E_{0}$. In the rest of the section, we formally show how this task can be done in a polynomial time, with respect to the system size $N$ and the variance $\varL$.
The algorithm informally proceeds as follows:
\begin{enumerate}
    \item Start with a product state $\ket{\Psi}$, at the desired energy $E_{0}$. Define projectors $P_i$ on each site $i$ that annihilate the state $\ket{\Psi}$. The sum of these projectors $P = \sum_i P_i$ defines a Hamiltonian whose ground state is the product state. 
    \item Choose the width $\delta$ of the Lorentzian filter to be applied to $\ket{\Psi}$.
    \item Construct the parent Hamiltonian:
    \begin{align*}
        &\calH(E_{0}, \delta) =\sum_i \calF_{\calL}^\dag P_i \calF_{\calL},\\
        &\text{where} \quad \calF_{\calL} = (1+i\dinv(H-E_{0})).
    \end{align*}
    The parent Hamiltonian $\calH$ has a unique ground state that corresponds to the filtered state and it remains gapped (see Sec.~\ref{sec:Algorithm}).
    \item Prepare the filtered state $\ket{\Phi}$ via adiabatic evolution of the product state 
    $\ket{\Psi}$ with $\calH(\delta)$ by slowly increasing $\dinv (s) = s \dinv$ from $s = 0$ to $1$. In Appendix \ref{PolyRuntimeProof}, we show that the adiabatic evolution time is $T = \poly(N, \dinv)$. Note that $\calH (s = 0) = P$. In App.~\ref{TrotterDepth}, we address the circuit depth required for Trotterized time evolution.
\end{enumerate}
For typical product states and a typical (local) Hamiltonian, the energy variance of the filtered state decreases as $\mathcal{O}(\delta \sqrt{N})$ in the limit of small $\delta$. By choosing a suitable $\delta$, we can thus prepare a filtered product state of arbitrary energy variance. 

\subsection{The algorithm}\label{sec:Algorithm}

Let $H$ be an arbitrary local Hamiltonian on $N$ sites with eigenbasis $H\ket{e_n} = e_n \ket{e_n}$ and
$\ket{\Psi}$ be a product state for which we can define local projectors $P_i$ that annihilate it. Let the energy of the product state be $E_{0}$. We set out to prepare the filtered state $\ket{\Phi}$, corresponding to filter $\calF_{\calL}$. This state is given by $\ket{\Phi} \propto \calF^{-1}_{\calL
}\ket{\Psi}$. In this paper, we will use the Lorentzian filter $\calF_{\calL}$, but the strategy works for any filter that is polynomial in $H$. The Lorentzian filter is given by 

\be \label{LorentzianF}
    \calF_{\calL}(E, \delta) = (1+i\dinv(H-E)),
\ee
with the corresponding filtered state :
\begin{align}
    &\ket{\Phi} \propto \calF^{-1}_{\calL}\ket{\Psi} = \sum_{n} \frac{c_n}{1+i\dinv(e_n-E)} \ket{e_n},\\
    &\abs{\bra{e_n}\ket{\Phi}}^2 \propto \frac{\abs{c_n}^2}{1+\delta^{-2}(e_n-E)^2}. \label{LorentzianSuppression}
\end{align}
Note that the probability amplitude of $\ket{e_n}$ has been suppressed here by a Lorentzian factor 
$\frac{1}{1+\delta^{-2}(e_n-E)^2}$.

\paragraph*{Parent Hamiltonian:} For a filter $\calF_{\calL}$ and initial state $\ket{\Psi}$ we can define the parent Hamiltonian $\calH$:
\be
\calH = \sum_i \calF^\dag_{\calL} P_i \calF_{\calL}.
\ee
Note that for a local Hamiltonian $H$, the operator norm of the parent Hamiltonian scales as $\norm{\calH} = \calO(N^3\delta^{-2})$. We take this scaling into account when establishing the bounds for the runtime and circuit depth. To show that $\ket{\Phi}$ is the ground state of $\calH$ note that by construction, $\ket{\Psi}$ is the unique ground state of $\sum_{i=1}^{N}P_i$, with zero energy.
Note that $\calH$ is a sum of positive semi-definite terms since:
    \be
    \calF^\dag P_i \calF = \calF^\dag P_i P_i \calF = r_i^\dag r_i \quad \text{where} \quad r_i \equiv P_i \calF,
    \ee
    thus the ground state of the parent Hamiltonian $\mathcal{H}$ has energy $E\geq 0$. Furthermore, note that, by construction, 
    \be 
    P_i \ket{\Psi} = P_i \calF \calF^{-1} \ket{\Psi} = \calF^\dag P_i \calF \calF^{-1} \ket{\Psi} = 0.
    \ee
    Hence the parent Hamiltonian $\calH$ has a unique ground state that corresponds to the filtered state $\ket{\Phi}$
    \be
    \ket{\mathrm{GS}}_\calH = \frac{1}{\calN} \calF^{-1} \ket{\Psi}= \ket{\Phi},
    \ee
    with ground state energy 0 and normalization factor $\calN$.
   
\paragraph*{Gap of the Parent Hamiltonian $\calH$.---}To ensure that we can efficiently perform the adiabatic evolution, we need to ensure that the parent Hamiltonian $\calH$ is gapped. We show this using the martingale method \cite{koma1995spectral}; in particular, we show that $\calH^2-\Delta \calH \succeq 0$ for a gap of $\Delta = 1$. The proof is given in Appendix \ref{gapApp}.

\paragraph*{Adiabatic runtime.---}Firstly, note that the parent Hamiltonian $\calH(\dinv = 0) = \sumP$, which is the sum of projectors with the initial product state $\ket{\Psi}$ as the ground state. Secondly, note that $\calH$ remains gapped for any finite $\dinv$ value. Thus, we can construct a gapped adiabatic path from the initial product state $\ket{\Psi}$ to the filtered state $\ket{\Phi}$ by starting with $\calH(\dinv = 0)$ and increasing $\dinv$ from 0 to the desired filter width. In the Appendix \ref{PolyRuntimeProof}, we look at the adiabatic runtime in more detail and establish that the filtered state $\ket{\Phi}$ can be prepared on a quantum computer with an adiabatic evolution time $T = \poly(N, \delta^{-1})$ and consequently with circuit depth $D = \poly(N, \delta^{-1})$.

\subsection{Considerations about product states}

Given a local Hamiltonian $H = \sum_j h_j$, the energy variance of a typical product state $\ket{\Psi}= \otimes_i \ket{\phi_i}$ scales as
\begin{align}
&\sigma^2_{0} = \bra{\Psi}H^2\ket{\Psi} - \bra{\Psi}H\ket{\Psi}^2 = \calO(N).
\end{align}
Both the mean energy $E_{0} = \bra{\Psi}H\ket{\Psi}$ and the variance $\sigma^2_{0}$ can be efficiently computed classically, and the projectors $P_i$ can be constructed as $P_i = \Id - \ket{\phi_i}\bra{\phi_i}$. 
Note also that for a local $H$, product states span an extensive energy range \cite{lieb1973classical}.

\paragraph*{Energy distribution of product states.---}The result of applying the filter $\calF$ on the state $\ket{\Psi}$ depends on the initial energy distribution, $\abs{c_n}^2$, given by the coefficients in the energy eigenbasis,
\begin{equation}
    \ket{\Psi} = \sum c_n \ket{e_n}.
\end{equation}
For a product state and a local Hamiltonian, this distribution can, to some extent, be approximated by a Gaussian in the thermodynamic limit, as shown in~\cite{hartmann2004gaussian,keating2015} using the central limit theorem.
In \cite{rai2023matrix}, authors extend this result using Berry-Essen theorem to show that the difference in the cumulative distribution of a product state and a Gaussian scales with the system size as $\mathcal{O}(N^{-1/2})$ in the worst case scenario. An example of the worst-case scaling is given by a state $\ket{\Psi} = \ket{+}^{\otimes N}$ for a non-interacting $H = \sum_i Z_i$ which has a binomial density of states. Thus, there are initial states, for which the lowest variance we can possibly reach with this procedure is $\mathcal{O}(1)$. However, for typical, interacting systems, this difference in distribution is much smaller.

\paragraph*{Energy variance of the filtered state.---}In order to estimate the dependence of the filtered state energy variance with the width of the Lorentzian filter $\delta$, we use a Gaussian approximation for the energy distribution of the initial product state, justified, in the generic case, by the above considerations. More concretely, we will assume that, for a typical product state, the energy distribution is exactly Gaussian, with mean $E_0$ and variance $\sigma_0$. Based on this approximation, in Appendix \ref{TheoryDecay}, we show that the energy variance of the filtered state scales as

\begin{equation}
    \sigma^2_{\calL}/\sigma_{0} \approx \delta \sqrt{2/\pi} \quad \text{for} \quad \delta/\sqrt{2\sigma^2_0}\ll 1.
\end{equation}
Note that in Appendix \ref{TheoryDecay} we only consider the scenario when the filter energy $E_F$ coincides with the product state energy $E_0$. When using a filter with $E_F \neq E_0$, one should not expect the energy of $\ket{\Phi}$ to be the filter parameter $E_F$ since the decay of the Gaussian eigenstate distribution is much faster than that of the Lorentzian filter, making the algorithm impractical in this case.

\section{Results}\label{sec:Results}

In the previous sections, we have shown that the Lorentzian filtering can be done in a polynomial time by adiabatic evolution since the gap of the parent Hamiltonian $\calH$ does not close throughout the evolution. However, the bounds on the adiabatic runtime are often loose, and in practice, one requires a much shorter runtime. We numerically benchmark the algorithm to show that we already observe a good agreement with theory for moderate system sizes $N$ and show that the algorithm requires a significantly shorter adiabatic evolution time. For the numerical results, we use the Transverse Field Ising (TFI) Hamiltonian $H$
\be \label{TFI}
    H = J\left (\sum_{i = 1}^{N-1} Z_i Z_{i+1} + \sum_{i=1}^{N} gX_i + hZ_i \right ),
\ee
 with coefficients $(J, g, h) = (1, -1.05, 0.5)$, far away from integrability. The $(X_i, Z_i)$ denotes the respective Pauli matrix on site $i$. We consider two types of initial product states $\ket{\Psi}$. Firstly, the antiferromagnetic (AFM) state $\ket{\text{AFM}} = \ket{1 0 1 \dots}$. Secondly, we look at a family of translationally invariant product states of the form $\ket{p(\theta)} = (\cos(\theta) \ket{0}+\sin(\theta)\ket{1})^{\otimes N}$. In principle, varying $\theta$ allows us to cover an extensive range of energies (see Appendix \ref{TI_PS}). While in the thermodynamic limit, the product states have a Gaussian eigenstate distribution, for small systems, the distribution can differ. We investigate the case for $\theta = \pi/6$ as the state lies closer to the center of the spectrum, where the density of states is larger, and has an eigenstate distribution closer to the Gaussian one. We expect this to give a better agreement with the theory for small systems.
 
 For a given product state $\ket{\Psi}$, we create the projectors $P_i$ on each site that annihilate it. We proceed to investigate the parent Hamiltonian:
\be
\label{Hp}
\calH =(1- \frac{i}{\delta}(H-E_F))\left ( \sum_{i=1}^N P_i \right )(1+ \frac{i}{\delta}(H-E_F)).
\ee

In this section, we show numerical results obtained with exact diagonalization for system sizes up to $N=18$ sites. 
Reaching larger system sizes with a classical simulation proves to be a computationally prohibitive task. In particular, since we are targeting states in the middle of the spectrum, the entanglement entropy is expected to scale as a volume law, preventing a systematic exploration with matrix product state methods~\cite{Schollwoeck_2011}, as we discuss more in detail in Appendix \ref{MPSAppendix}.

\subsection{Expected decrease of the energy variance}

\begin{figure}
    \includegraphics[width = \linewidth]{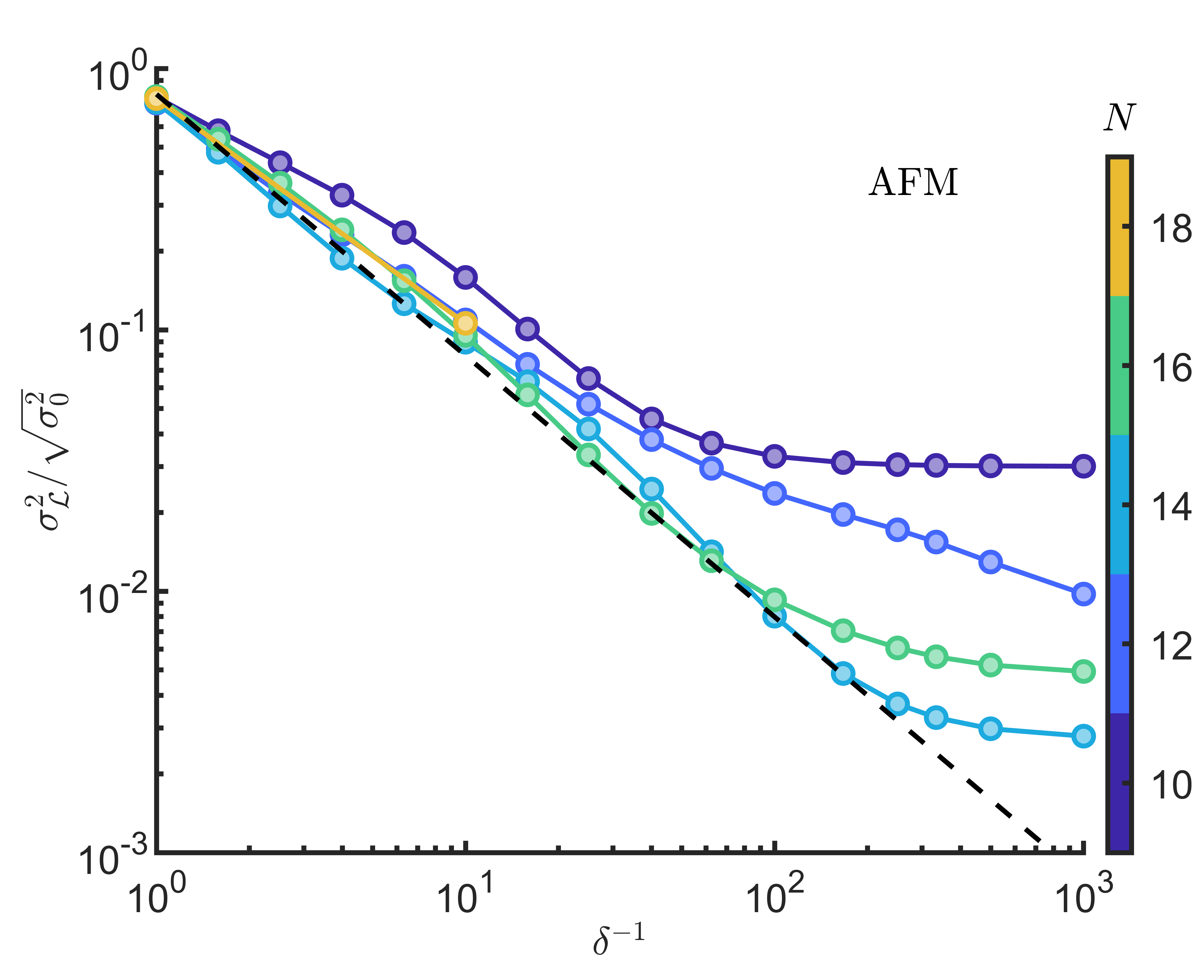}
    \caption{\label{TheoPlotAFM} The figure shows, for various system sizes $N$, the performance of the filter applied on the AFM product state at the mean energy which, in the thermodynamic limit, corresponds to energy density $E_{0}/JN = 1$. The black line shows the expected variance decay (Eq.~\ref{TheoVarTrendline}).}  
\end{figure}

\begin{figure}    
    \includegraphics[width = \linewidth]{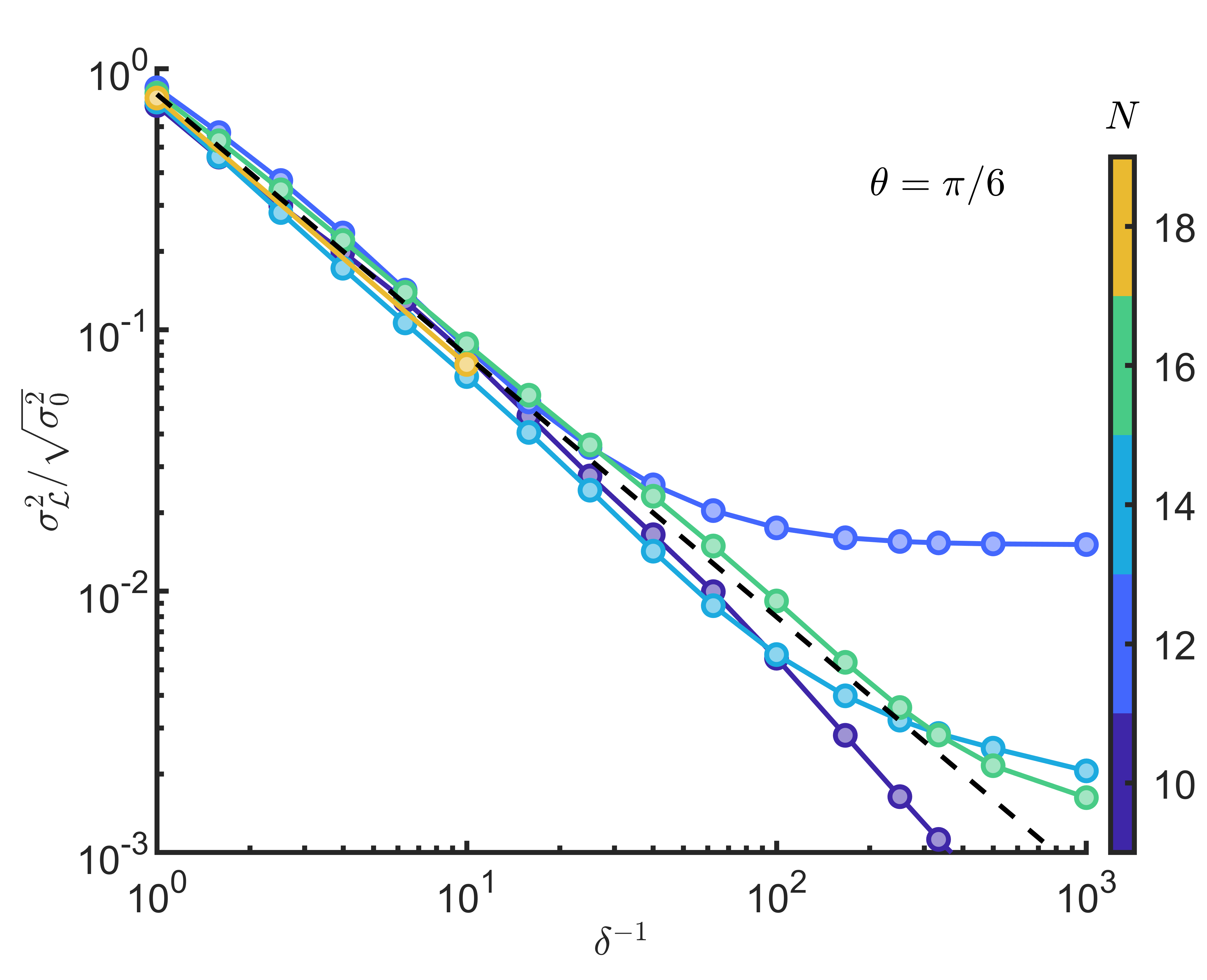}
    \caption{\label{TheoPlotPi6} The figure shows, for various system sizes $N$, the performance of the filter applied on the $\ket{p(\theta = \pi/6)}$ product state at the mean energy which, in the thermodynamic limit, corresponds to energy density $E_0/JN = -0.409$. The black line shows the expected variance decay (Eq.~\ref{TheoVarTrendline}).}  
\end{figure}

Here, we numerically investigate the ground state of $\calH$ corresponding to the filtered product state. In particular, we investigate how the variance $\varL$ of the Lorentzian-filtered state depends on the parameter $\delta$ for various system sizes $N$. In Appendix \ref{TheoryDecay}, we derive the relationship between variance and $\delta$. For low $\delta$ values, this relationship reduces to:
\begin{equation}\label{TheoVarTrendline}
    \sigma^2_{\calL}/\sqrt{\sigma^2_{0}} = \delta \sqrt{2/\pi}.
\end{equation}
In Fig.~\ref{TheoPlotAFM}, we consider the decay of the variance for the AFM initial state, which lies at the edge of the spectrum of the Hamiltonian $H$. Fig.~\ref{TheoPlotPi6} correspondingly shows the results for the initial state $\ket{p(\theta = \pi/6)}$, which lies more toward the center of the spectrum. 
We observe a good agreement with theory in both cases, with a better agreement for the small system sizes in the case of $\ket{p(\theta = \pi/6)}$ initial state.

\subsection{Adiabatic Evolution}

 To better understand the actual adiabatic runtime $T$ required, we numerically investigate the performance of the adiabatic evolution in practice. On a quantum device, one could prepare the filtered state by time evolving the product state with $\calH(\delta)$ while smoothly increasing $\delta^{-1}$ from 0 to the desired value. In particular, for this study, we choose the adiabatic scheduling $\delta^{-1}(s) = \sin(\pi/2 \sin(s\pi/2)^2)^2 \delta^{-1}_{\max}$ which ensures a slower change of the parameter at the beginning and end of the evolution. This scheduling of the adiabatic evolution is essential for an optimal runtime \cite{reichardt2004quantum}.
 
\begin{figure}[ht]
    \includegraphics[width = \linewidth]{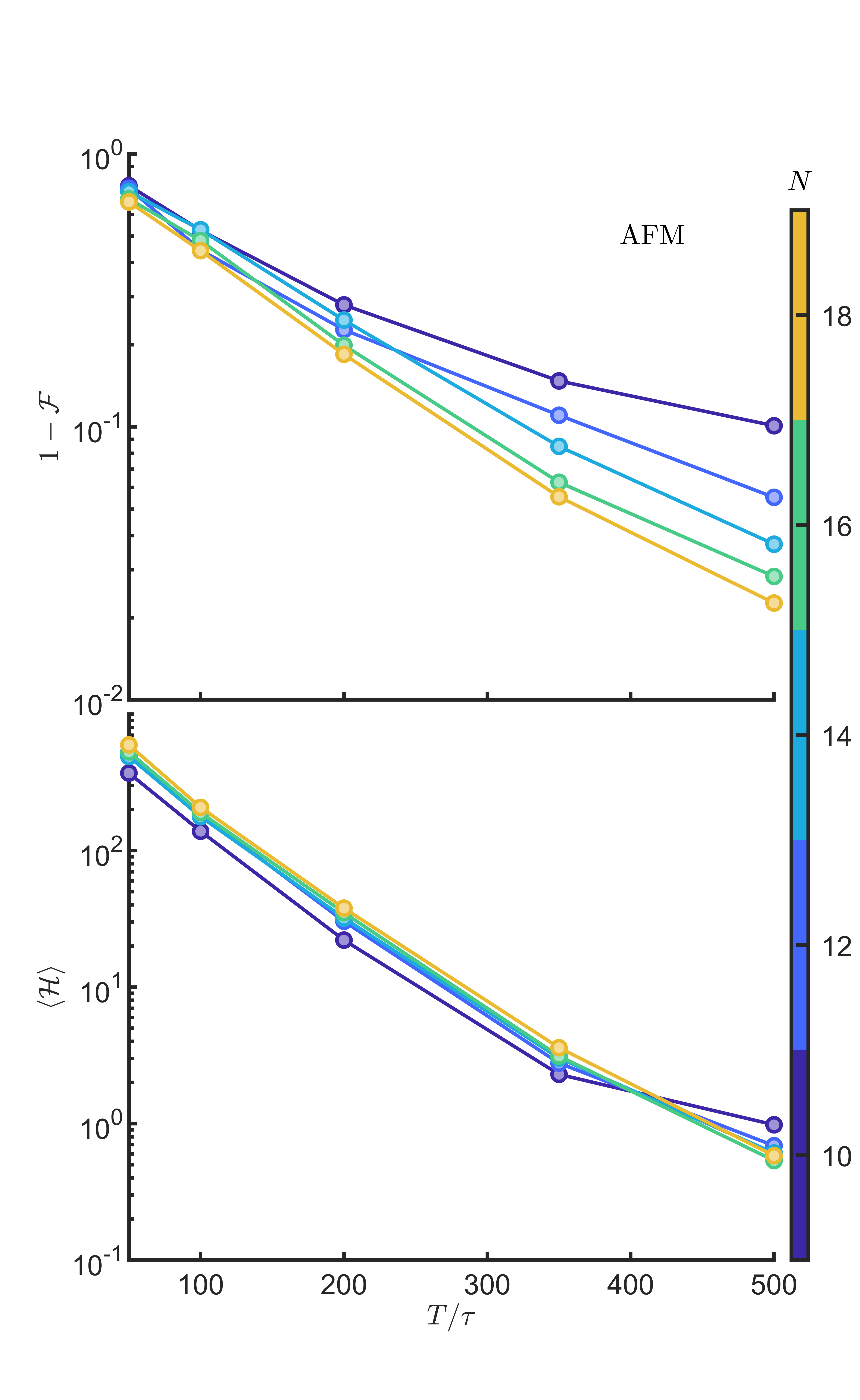}
    \caption{\label{AdiPlotAFM} The figure shows results for adiabatic evolution of the $\ket{AFM}$ initial product state for final filter width $\delta = 0.1$ and several system sizes $N$.  
    We use $\tau = 0.1$. 
    The upper plot shows one minus the fidelity between the evolved state and the exact filtered one as a function of the number of timesteps $T/\tau$.
    The lower plot shows the parent Hamiltonian $\calH$ energy at the end of the adiabatic evolution.}
\end{figure}

\begin{figure}[ht]   
    \includegraphics[width = \linewidth]{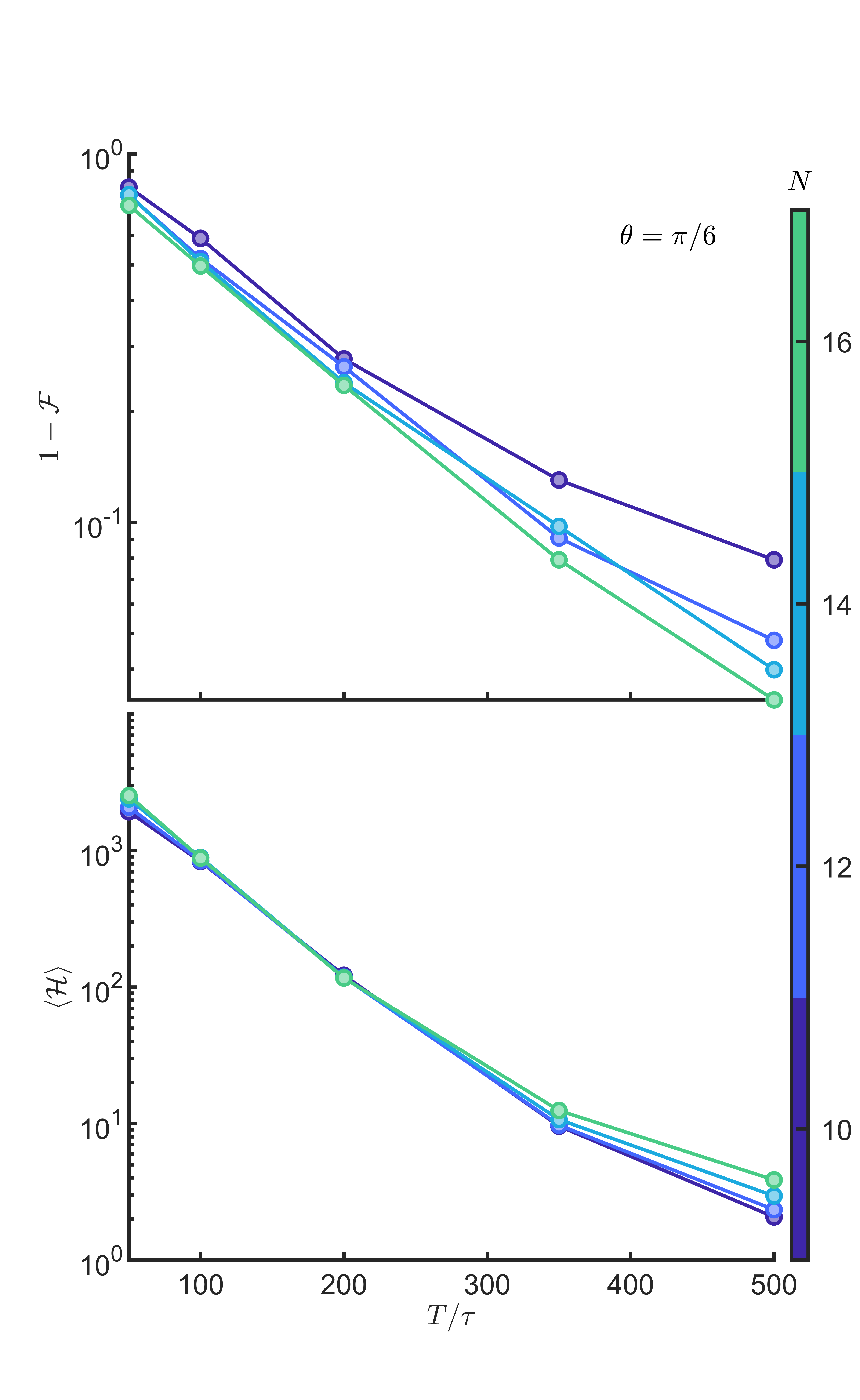}
    \caption{\label{AdiPlotPi6} The figure shows results for adiabatic evolution of $\ket{p(\theta = \pi/6)}$ initial product state for final filter width $\delta = 0.1$ and various system sizes $N$. We use $\tau = 0.1$. The upper plot shows one minus the fidelity between the evolved state and the exact filtered one as a function of the number of timesteps $T/\tau$.
    The lower plot shows the parent Hamiltonian $\calH$ energy at the end of the adiabatic evolution.}  
\end{figure}

To investigate the accuracy of the adiabatic evolution for various system sizes, we study the fidelity of the adiabatic state $\calF$ with respect to the exact filtered state and monitor the ground state energy of the parent Hamiltonian $\calH$ since, for perfectly adiabatic evolution, it should remain zero throughout. Note that the parent Hamiltonian $\calH$ (Eq.~\ref{Hp}) has a norm that increases with $\dinv$. To keep it normalized throughout adiabatic evolution, we rescale it to $\tilde{\calH}(\delta) = \calH(\delta)/(1+\delta^{-2})$. Note that the rescaling affects the gap, which we take into account when establishing the runtime. We apply the adiabatic evolution in the following way:
\begin{equation}
    \ket{\Phi}_{\mathrm{adi}} = \prod_{l = 1}^{T/\tau} \exp(-i \tau \tilde{\calH} \{\delta(l\frac{\tau}{T})\}) \ket{\Psi},
\end{equation}
where $\tau$ is the discrete adiabatic time step, and $T$ is the total adiabatic evolution time. To benchmark the adiabatic evolution, we fix the desired filter width to be $\delta = 0.1$ and perform the evolution for several system sizes $N$ and runtimes $T$. In Fig.~\ref{AdiPlotAFM} and Fig.~\ref{AdiPlotPi6}, we show the dependence of the fidelity $\calF$ and the parent Hamiltonian $\calH$ energy with respect to the adiabatic evolution time $T$. In both cases, the parent Hamiltonian $\calH$ energy appears independent of the system size, while the fidelity shows a favorable behavior towards larger $N$. These results provide a more favorable scaling for the adiabatic runtime $T$ compared to the adiabatic theorem, which suggests that $T = \calO(N^3)$. The results suggest that the adiabatic evolution runtime could be taken independently of the system size to prepare the state with a given fidelity $\calF$. However, in Appendix~\ref{TrotterDepth}, we show that the circuit depth required to implement this evolution via the first-order Trotterization scales as $\calO(N^2)$ since $\norm{\calH} = \calO(N^3)$.

\section{Discussion}\label{sec:Discussion}

In this section, we review the results from Section \ref{sec:Results} and discuss the behavior and drawbacks of the Lorentzian filtering for finite system sizes. 

Firstly, Fig.~\ref{TheoPlotAFM} and Fig.~\ref{TheoPlotPi6} show that, even for small system sizes $N$, the variance of the filtered state follows the decreasing behavior expected in the thermodynamic limit (Eq.~\ref{TheoVarTrendline}). In the numerics, we observe that the smaller systems start to deviate from this behavior first. The first reason for this behavior is that the eigenstate distribution of the product state is not Gaussian for small systems. The second reason is the discreteness of the spectrum. In Eq.~\eqref{LorentzianSuppression}, we establish that the suppression due to Lorentzian filtering is given by $\frac{\abs{c_n}^2}{1+\delta^{-2}(e_n-E_0)^2}$. Suppose that the closest eigenstate is $\eta = \min_n{\abs{e_n - E}}$ away from the center of the filter. It is clear that further lowering $\delta$ leads to no further filtering when $\dinv \eta \gtrsim 1$. However, for larger systems, the discreteness of the spectrum should not pose any problems. Note that for a Hamiltonian $H$ on $N$ sites, the energy extent is $O(N)$, while there are $2^N$ eigenstates. This means that the density of states, along with the number of states supported by the product state, increases exponentially with $N$, and thus $\eta$ becomes exponentially small in $N$. 

Secondly, in Fig.~\ref{AdiPlotAFM} and Fig.~\ref{AdiPlotPi6}, we observe that the adiabatic runtime required to reach a certain precision is essentially independent of the system size $N$. 
Even though the fidelity is only accessible in exact calculations, 
measuring the energy of the parent Hamiltonian $\calH$ at the end of the evolution,
as seen in the figures, provides a reliable indication that the adiabatic evolution has indeed prepared the required state.
The numerics indicate that the runtime of the algorithm is much faster than expected, and one can prepare a state of given $\delta$ with a fixed number of Trotter steps that are independent of $N$. 
Thirdly, perhaps the biggest drawback of the proposed algorithm is the fact that it requires evolving the system under a geometrically non-local Hamiltonian.
Whereas this would, in principle, be possible on some platforms, like trapped ions \cite{Grzesiak_2020}, on other quantum computer implementations with only local gates available, it would require $\calO(N)$ additional swap operations, which will increase the cost of the algorithm and, potentially, render it unsuitable for NISQ devices.
This is, however, not surprising when considering that the states we are trying to filter are approximations of excited eigenstates of the Hamiltonian, which usually exhibit an entanglement volume law~\cite{PRXQuantum.3.030201, HUANG2021115373, Anshu_2022}, and thus cannot be encoded, in general, in the ground state of a geometrically local Hamiltonian (which would only accommodate area law entanglement). 
\section{Conclusion}\label{sec:Conclusion}
We have developed a novel algorithm for preparing states with an arbitrarily small energy variance at a target energy on a quantum computer. 
The algorithm effectively prepares a state that would result from applying a Lorentzian filter of width $\delta$ to an initial product state. The filtered state is, in fact, adiabatically prepared as the ground state of a parent Hamiltonian. Furthermore, we prove that the algorithm is efficient since the parent Hamiltonian remains gapped with a gap independent of system size; thus, it can be efficiently prepared with adiabatic evolution. Indeed, the running time of the algorithm is polynomial both in the system size and the inverse width.

We show that the variance of the filtered state decreases with the parameter $\delta$ that defines the Lorentzian filter, and already for moderate system sizes, we observe good agreement with the asymptotic predictions for large sizes.

While according to the theoretical bounds, the adiabatic runtime scales as $O(N^3\delta^{-4})$, our numerics suggest that, in practice, the required time is much more favorable due to the looseness of the theoretical bound from the adiabatic theorem. In practice, we estimate that the adiabatic runtime scales as $O(\delta^{-4})$, thus allowing us to perform adiabatic evolution and prepare the finite energy state with circuit depth $O(N^2\delta^{-4})$ (see Appendix \ref{TrotterDepth}). 

Our approach provides a new way of probing finite energy physics on quantum devices by directly giving access to small energy variance states. Having access to the state itself provides novel ways to probe finite energy regimes of isolated quantum systems and allows this algorithm to serve as a subroutine in more complicated analysis on quantum devices. Future work should focus on developing new algorithms that take advantage of the access to filtered product states.

\acknowledgements{We thank Álvaro M. Alhambra and Georgios Styliaris for helpful discussions. We
acknowledge the support from the German Federal Ministry of Education and Research (BMBF) through FermiQP (Grant No. 13N15890) and EQUAHUMO (Grant No. 13N16066) within the funding program quantum technologies—from basic research to market. This research is
part of the Munich Quantum Valley (MQV), which is supported by the Bavarian state government with funds from the Hightech Agenda Bayern Plus.
This work was partially supported by the Deutsche Forschungsgemeinschaft (DFG, German Research Foundation) under Germany's Excellence Strategy -- EXC-2111 -- 390814868;  
and by the EU-QUANTERA project TNiSQ (BA 6059/1-1).}

\bibliographystyle{quantum}
\bibliography{main}

\appendix
\section{Parent Hamiltonin Gap} \label{gapApp}

In this appendix, we prove that the parent Hamiltonian $\calH$ is gapped with a gap $\Delta \geq 1$. Consider the parent Hamiltonian $\calH$:
\begin{align}
    \calF &= (1+i\dinv(H-E)),\\
    \calH &= \calF^\dagger \sumP \calF,
\end{align}
and note that $\calF^\dagger \calF \succeq 1$. In fact, this is the only requirement for the filter that we use in the proof, so in principle, the results also hold for other, higher-order filters that satisfy this property.
We use the martingale method \cite{koma1995spectral} to prove that  $\calH$ is gapped; in particular, we show that $\calH^2-\Delta \calH \succeq 0$ for a gap of $\Delta = 1$.
\begin{align*}
    &\calH^2-\calH =\\ &=\calF^\dagger \sumP  \calF^\dagger \calF  \sumP \calF - \calF^\dagger \sumP  \calF\\
    &= \calF^\dagger \sumP  \calF^\dagger \calF  \sumP \calF - \calF^\dagger \sumPsq  \calF\\
    &\succeq \calF^\dagger \sumP  \calF^\dagger \calF  \sumP \calF - \calF^\dagger \sumP ^2\calF\\
    &= \calF^\dagger \sumP ( \calF^\dagger \calF -1) \sumP \calF\\
    &= V^\dagger ( \calF^\dagger \calF -1) V\\
    &\succeq 0,
\end{align*}
since  $(\calF^\dagger \calF -1) \succeq 0$, where $V = \sumP \calF$. In line 3, we have used the fact that $P_i^2 = P_i$, and in line 4, we have relied on the fact that any term $P_iP_j\succeq 0$, since each of the projectors $P_i \succeq 0$ and they commute $\comm{P_i}{P_j}\succeq 0$.

\section{Adiabatic runtime} \label{PolyRuntimeProof}

In this appendix, we discuss the adiabatic runtime required for filtering. The adiabatic theorem states that if the adiabatic evolution is slow enough and the state is gapped then the system will remain in the instantaneous ground state of the Hamiltonian $H(s)$. The result from \cite{Kato1950, reichardt2004quantum, Amin_2009} states that the required runtime for the adiabatic evolution is:
\be
    T \geq \max_s (\norm{\dot{H}(s)}/\Delta(s)^2),
\ee
where we smoothly vary the parameter $s$ from $0$ to $1$. To keep the norm of the parent Hamiltonian independent of $\dinv$ throughout the adiabatic evolution, we rescale it as follows:
\begin{align*}
    \tilde{\calH}(s) &= \frac{1}{1+s^2\delta^{-2}} V\sumP V^{\dag}, \quad \text{where}\\
    V&= \left( 1+ is\dinv (H-E) \right).
\end{align*}
When rescaled in this way, the gap bound from Appendix \ref{gapApp} states that the gap $\Delta(s) \geq \frac{1}{1+s^2\delta^{-2}}$. Furthermore, by differentiating $\tilde{\calH}(s)$ with respect to $s$ we obtain that:
\begin{align*}
    \dot{\tilde{\calH}}(s) &= \frac{-2s\delta^{-2}}{(1+s^2\delta^{-2})^2}V\sumP V^{\dag}\\&+\frac{1}{1+s^2\delta^{-2}}(\dot{V}\sumP V^\dag+V\sumP\dot{V}^{\dag})\\
    &= \frac{1}{1+s^2\delta^{-2}}(-2s\delta^{-2}\calH(s)\\&+\delta^{-1}(i(H-E)\sumP V^{\dagger} + h.c.)\\
    &= O(N^3),
\end{align*}
since $\norm{V} = \calO(\dinv N)$ and $\norm{H} = \calO(N)$
Combining the previous bound for $\norm{\dot{\tilde{\calH}}(s)}$ and the scaling of the gap $\Delta(s)$ we arrive at the adiabatic runtime:
\begin{align*}
    T &\geq \max_s (\norm{\dot{\tilde{\calH}}(s)}/\Delta(s)^2)\\
    &= O(N^3\delta^{-4}),
\end{align*}
which gives the $T = \poly(N,\delta)$ adiabatic runtime. 

Secondly, we use the fact that the time evolution on quantum computers can be performed efficiently for a local Hamiltonian $H = \sum_{\gamma = 1}^{M} h_{\gamma}$ using any of the algorithms \cite{Low2017, Low_2019, Childs2021}. Thus, the evolution can be simulated with a $\poly(N,\delta)$-depth quantum circuit. In Appendix \ref{TrotterDepth}, we explicitly show the depth required to implement the first-order Trotter evolution.

\section{Trotter Circuit depth for the adiabatic evolution}
\label{TrotterDepth}

This appendix shows that the first-order Trotterized time evolution of $\calH$ can be implemented with depth $D = O(TN^2)$. Suppose $\{P_i\}$ are a set of one-site projectors, and $H = \sum_{\gamma} ^M h_{\gamma}$ is a local Hamiltonian such the range of interactions $\abs{\supp{h_{\gamma}}}\leq w$ and that for any site $i$ there are at most $v$ terms $h_{\gamma}$ acting on it.   
Then the first-order Trotterized evolution of the parent Hamiltonian $\calH = (1-i\dinv(H-E_0))\sumP(1+i\dinv (H-E_0))$ for time $T$ with timestep $\tau$ can be implemented with circuit depth $D = \mathcal{O}(TN^2)$

We prove this as follows. Let's consider the circuit depth required to implement a single Trotter step of the parent Hamiltonian $\calH$ for a small $\tau$.
 The parent Hamiltonian $\calH$ can be split into parts: 
\begin{align*}
    \calH &= \sumP\\ &-\frac{i}{\delta}\comm{H}{\sumP}\nonumber \\&+ \frac{1}{\delta^2}(H-E_0)\sumP(H-E_0).
\end{align*}
It's clear that the terms in $\sumP$ are 1-local; thus, their evolution can be implemented with a $D = O(1)$ circuit. Similarly, for the second set of terms $\comm{H}{\sumP}$, it's clear to see that it is a sum of $\calO(N)$ geometrically local terms of weight at most $w$; thus can be implemented with a circuit of depth $O(1)$. The most complicated terms arise from the third set of terms $(H-E_0)\sumP(H-E_0)$. First, note that we can implement any evolution of type $\exp(-i\tau h_iP_jh_k)$ with a constant circuit depth, assuming that the qubits have all-to-all connectivity. Suppose we pick any three sites $i,j,k$. Then, the support of all the terms acting on all 3 of these vertices can be described by: 
\begin{align*}
    S_{ijk} &= \{i-(w-1)\dots i+(w-1) \\&\cup j-(w-1)\dots j+(w-1)\\ &\cup k-(w-1)\dots k+(w-1)\},
\end{align*}
and $\abs{S_{ijk}} \leq 3(2w-1)$. Since for any vertex $i$, there are at most $v$ terms in $H$ and one term in $\sumP$ acting on it, the number of terms acting on all 3 of these vertices is at most $(v+1)^3$,
and they all have support within $S_{ijk}$, allowing them to be implemented with circuit depth independent of $N$. There are $N^3$ combinations of vertices $i,j,k$, but each of these terms only have support on sites within $S_{ijk}$; thus, we can implement the evolution of $N/(3(2w-1))$
combinations in parallel, leading to the necessary circuit depth being $O(N^2)$. Thus, the entire evolution of the parent Hamiltonian $\calH$ can be implemented with circuit depth $D = O(T/\tau N^2) = O(TN^2)$, which concludes the proof. 

\section{Filtered energy variance}\label{TheoryDecay}

In this appendix, we derive the relationship between the filter width $\delta$ and the energy variance of $\ket{\Phi}$ by approximating that the eigenstate distribution of the initial product state $\ket{\Psi}$ is Gaussian, which has been shown to be a good approximation for local Hamiltonians at large $N$ ~\cite{hartmann2004gaussian, anshu2016concentration, kuwahara2016}. Assume $\ket{\Psi}$ has a Gaussian distribution of eigenstates $\calG(E_{0}, \sigma^2_0)$ centered at $E_{0} = \bra{\Psi}H\ket{\Psi}$ and with variance $\sigma^2_0 = \bra{\Psi}H^2\ket{\Psi} - \bra{\Psi}H\ket{\Psi}^2$. This allows us to estimate the variance upon the application of the filter $\calL(E_F, \delta)$ as follows:
\begin{align*}
    \varL &= \frac{\int_{-\infty}^{\infty} e^2\calG(E_{0}, \sigma^2_0)\calL(E_F, \delta)de }{\int_{-\infty}^{\infty}\calG(E_{0}, \sigma^2_0))\calL(E_F, \delta)de}\\
    &-\left(\frac{\int_{-\infty}^{\infty} e\calG(E_{0}, \sigma^2_0)\calL(E_F, \delta)de }{\int_{-\infty}^{\infty}\calG(E_{0}, \sigma^2_0))\calL(E_F, \delta)de}\right)^2\\
    &= \frac{\calI_2}{\calI_0}-\left(\frac{\calI_1}{\calI_0}\right)^2.
\end{align*}
We investigate the filtered energy variance when the Lorentzian filter is applied at the center of the product state energy, thus $E_F = E_{0}$. In this case, the above integrals can be evaluated to yield:
\begin{align*}
    \calI_0 &= \int_{-\infty}^{\infty}\frac{\exp(-(e-E_{0})^2/2\sigma^2_0)}{(e-E_{0})^2+\delta^2} de\\
    &= \int_{-\infty}^{\infty}\frac{\exp(-e'^2/2\sigma^2_0)}{e'^2+\delta^2} de'\\
    &= \frac{\pi} {\delta}\exp(\frac{\delta^2}{2\sigma^2_0}) \left[ 1-\erf\left(\delta/\sqrt{2\sigma^2_0}\right)\right ],\\
\end{align*}
\begin{align*}
    \calI_1 &= \int_{-\infty}^{\infty}\frac{\exp(-(e-E_{0})^2/2\sigma^2_0)}{(e-E_{0})^2+\delta^2} e de\\
    &= \int_{-\infty}^{\infty} \frac{\exp(-e'^2/2\sigma^2_0)}{e'^2+\delta^2} (e'+E_{0}) de' \\
    &= 0 + \int_{-\infty}^{\infty} \frac{\exp(-e'^2/2\sigma^2_0)}{e'^2+\delta^2} E_{0} de' \\
    &= \calI_0 E_{0},
\end{align*}
\begin{align*}
     \calI_2 &= \int_{-\infty}^{\infty}\frac{\exp(-(e-E_{0})^2/2\sigma^2_0)}{(e-E_{0})^2+\delta^2} e^2 de\\
    &= \int_{-\infty}^{\infty}\frac{\exp(-e'^2/2\sigma^2_0)}{e'^2+\delta^2} (e'+E_{0})^2 de'\\
    &= \int_{-\infty}^{\infty}\frac{\exp(-e'^2/2\sigma^2_0)}{e'^2+\delta^2} e'^2 de' + E^2_{0}\\
    &= -\pi \delta \exp(\frac{\delta^2}{2\sigma^2_0}) \left( 1-\erf(\delta/\sqrt{2\sigma^2_0})\right )\\ &+ \sqrt{2\pi\sigma^2_0}+E^2_{0}\\
    &= -\delta^2 \calI_0 + \sqrt{2\pi\sigma^2_0}+E^2_{0}.
\end{align*}
Combining the above results in the following filtered variance:
\begin{align}
    \varL &= -\delta^2 + \frac{\delta \exp(-\frac{\delta^2}{2\sigma^2_0})\sqrt{2\sigma^2_0/\pi}}{\left( 1-\erf(\delta/\sqrt{2\sigma^2_0})\right )}.
\end{align}
Note that in the limit where:
\begin{equation*}
    \lim_{\delta\to \infty}{\varL} = \sigma^2_0.
\end{equation*}
However, we are interested in the limit for small $\delta$ values $(\delta/\sqrt{2\sigma^2_0}<<1)$ for which we get the asymptotic behavior:
\begin{align}
    \varL &\approx \delta \sqrt{2\sigma^2_0/\pi} = \calO(\delta \sqrt{N}).
\end{align}

This establishes the bound for the scaling of the filtered energy variance.  

\section{Translationaly invariant product states}\label{TI_PS}

\begin{figure}
    \includegraphics{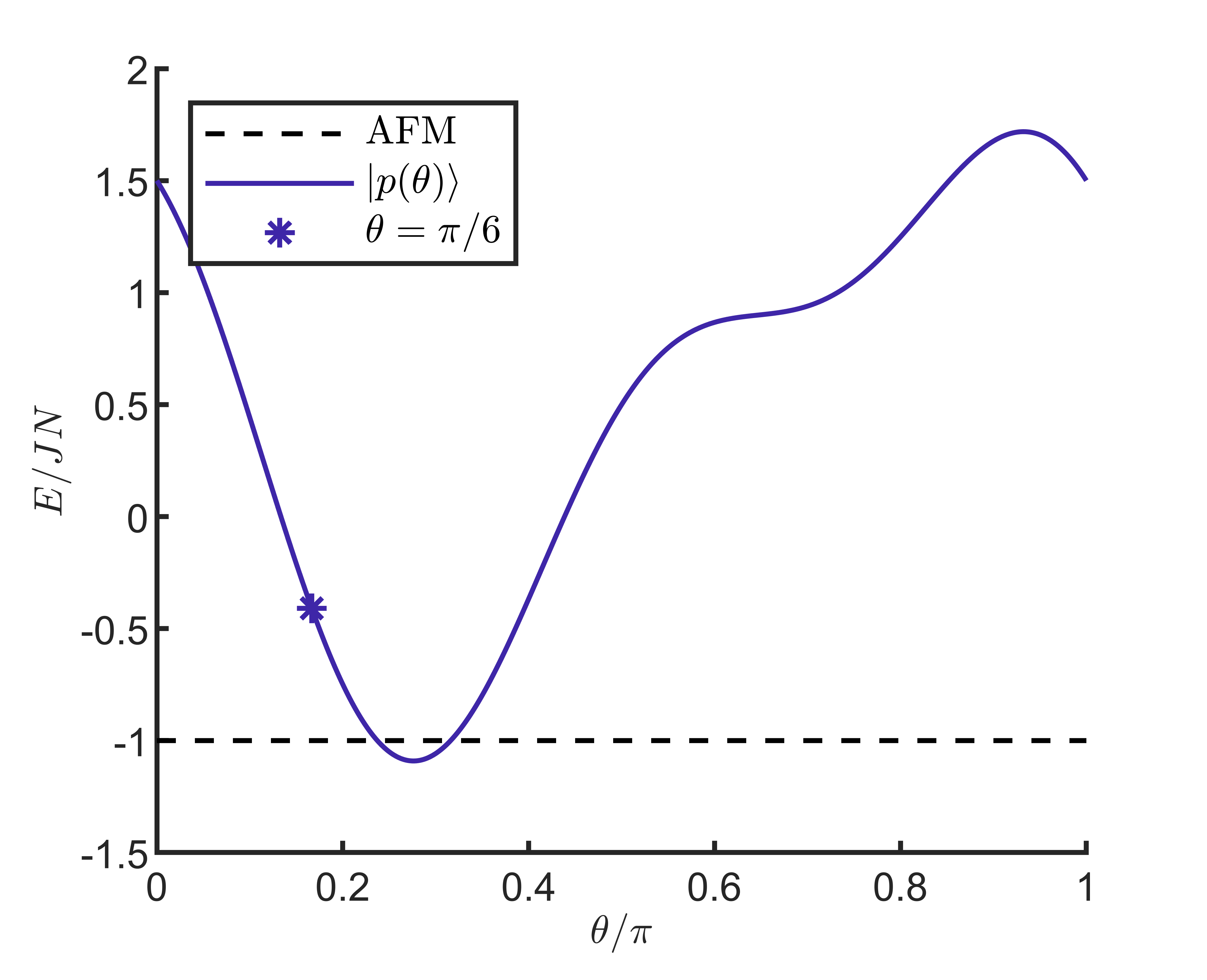}
    \caption{$E/JN$ dependence on the parameter $\theta$. We have marked the $\theta = \pi/6$ value used in the work. The AFM state has $E/JN = -1$ in the thermodynamic limit.}
    \label{fig:energyTheta}
\end{figure}

A family of states we consider in our investigation is translationally invariant product states, defined by:
\begin{equation}
    \ket{p(\theta)} = (\cos(\theta) \ket{0}+\sin(\theta)\ket{1})^{\otimes N}. 
\end{equation}
For the TFI Ising model Eq. ~\ref{TFI}, such product state parametrization covers a range of energies, allowing us to investigate the performance of the filter across the whole spectrum. In the thermodynamic limit, the energy of $\ket{p(\theta)}$ is $E/JN = \cos^2(2\theta)+h\cos(2\theta)+g\sin(2\theta)$. The dependence on the angle $\theta$ is depicted in Fig.~\ref{fig:energyTheta}.

\section{MPS investigation}\label{MPSAppendix}

We have limited the numerics in Section \ref{sec:Results} to only exact calculations for small systems. One could expect that matrix product state (MPS) methods~\cite{Schollwoeck_2011} would allow us to reach larger system sizes. However, this proves to be a very challenging task. In this appendix, we provide details about the drawbacks of such MPS calculations for this particular model and illustrate why it is hard to go beyond the system sizes accessible with the exact methods. 

\begin{figure}[H]
    \includegraphics{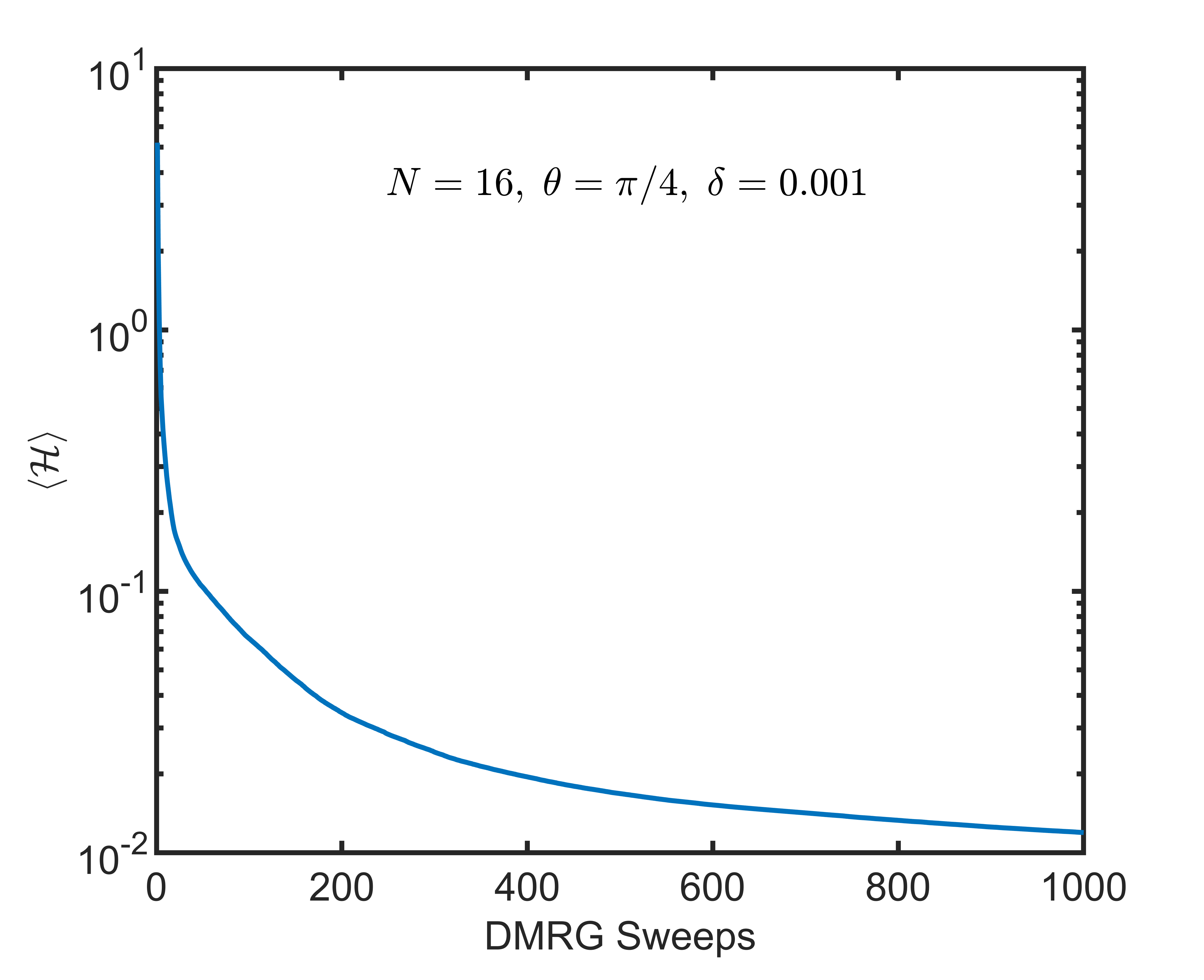}
    \caption{The figure shows the decay of the parent Hamiltonian $\calH$ energy with the number of DMRG sweeps. The convergence to the true value of $\expval{\calH} = 0$ happens slowly.}
    \label{DMRGFail}
\end{figure}

Firstly, we investigate the ground state of $\calH$ using DMRG. In Fig.~\ref{DMRGFail}, we show that we approach the ground state slowly in the number of sweeps. As shown in the figure, which illustrates the ground state search corresponding already for a small system size $N=16$ (which can be solved with exact diagonalization), even after 1000 sweeps, the error in the energy is considerably large (note that the exact value is zero).

\begin{figure}[H]
    \includegraphics{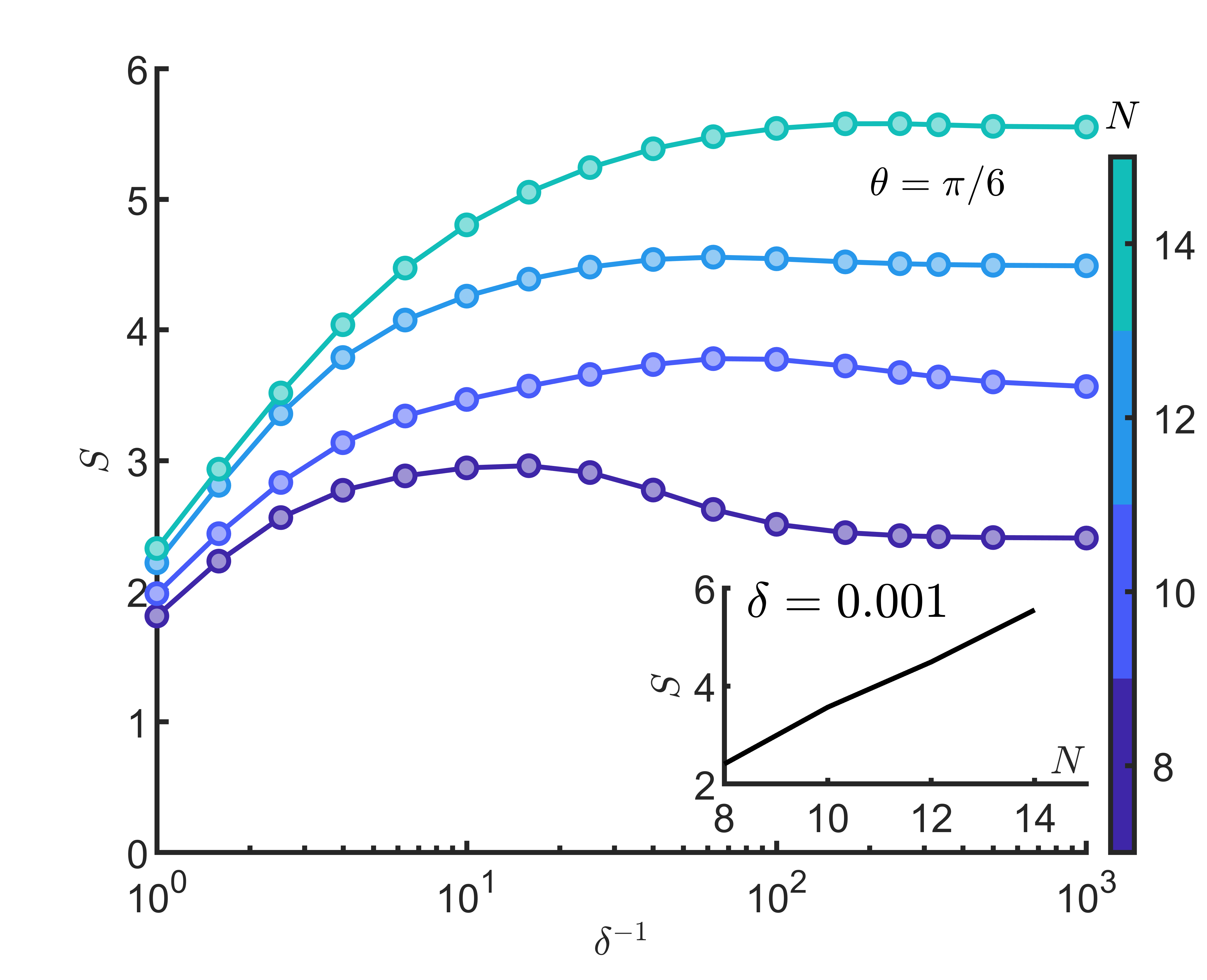}
    \caption{The figure shows the dependence of the bipartite entanglement entropy with $\dinv$ and for various system sizes $N$. We observe that for a given $\dinv$, the entanglement grows linearly with $N$.}
    \label{EntanglementEntropy}
\end{figure}

Secondly, we look at the entanglement entropy of the filtered states and how it depends on the system size $N$ and $\dinv$ to understand the fundamental limitations in their approximability as MPS. 
Fig.~\ref{EntanglementEntropy} shows the dependence of the entropy of half chain with $\dinv$ for various system sizes, obtained from our exact calculations. 
The results indicate a tendency to a linear increase with $N$ for sufficiently small $\delta$, consistent with a volume law.
We thus conclude that the MPS techniques will not be able to efficiently capture the ground state, respectively, the adiabatic evolution
of the parent Hamiltonian $\calH$. This behavior is expected since we are trying to approximate an energy eigenstate at the center of the spectrum, which follows the volume law of entanglement. The inability to express these states as MPS further solidifies the need for a quantum algorithm that is capable of preparing such states and investigating their properties. 
\linebreak

\end{document}